%% file: main.tex
\documentclass[conference]{IEEEtran}
\input{preamble.tex}

\begin{document}

\title{Hurry-up: Scaling Web Search on \\ Big/Little Multi-core Architectures}

\author[1]{Rajiv Nishtala}
\author[2]{Vinicius Petrucci}
\author[3]{Paul Carpenter}
\author[3]{Xavier Martorell} 
\affil[1]{Norwegian University of Science and Technology (rajiv.nishtala@ntnu.no)}
\affil[2]{Federal University Bahia, Brazil \& University of Pittsburgh, USA (vpetrucci@upitt.edu)}
\affil[3]{Barcelona Supercomputing Center, Spain (first.last@bsc.es)}

\maketitle

\begin{abstract}
Heterogeneous multi-core systems such as big/little architectures have been introduced as an attractive server design option with the potential to improve performance under power constraints in data centres. Since both big high-performing and little power-efficient cores can run on the same system sharing the workload processing, thread mapping/scheduling turns out to be much more challenging. This is particularly hard when considering the different trade-offs shaped by the heterogeneous cores on the quality-of-service (expressed as tail latency) experienced by user-facing applications, such as Web Search.

In this work, we present Hurry-up, a runtime thread mapping solution designed to select individual requests to run on the most appropriate heterogeneous cores to improve tail latency. Hurry-up accelerates compute-intensive requests on big cores, while letting less intensive threads to execute on little cores. We implement and deploy Hurry-up on a real 64-bit big/little architecture (ARM Juno), and show that, compared to a conservative policy on Linux, Hurry-up reduces the server tail latency by 39.5\% (mean).

\end{abstract}

\IEEEpeerreviewmaketitle
    
\input{introduction}
\input{motivation}
\input{algorithm}
\input{evaluation}
\input{relatedwork}
\input{conclusion}

\section*{Acknowledgement}

This work was funded by the European Union under grant agreement No 754337 (EuroEXA), the Brazilian federal government under CNPq grant (Process nº 430188/2018-8).

The experiments were conducted on the Juno board at the Barcelona Supercomputing Center, Spain.

\bibliographystyle{abbrv}
\bibliography{bibliography,defs}

\end{document}

%% file: preamble.tex
\newcommand{\ignore}[1]{}
\usepackage{fancyhdr}
\usepackage[normalem]{ulem}
\usepackage{textcomp}

\usepackage{pgfplots}
\usepackage{float}
\usepackage{fixltx2e}
\usepackage{graphicx}
\usepackage{amsmath}
\usepackage{algorithm}
\usepackage[noend]{algpseudocode}
\algrenewcommand\alglinenumber[1]{\footnotesize #1}
\usepackage{algorithmicx}

\newcommand*{\etal}{{et al.}}
\makeatletter
\algrenewcommand\ALG@beginalgorithmic{\footnotesize}
\makeatother
\makeatletter
\def\@copyrightspace{\relax}
\makeatother

\usepackage{enumitem}

\setlist[itemize]{noitemsep, topsep=3pt, leftmargin=*, align=left}
\setlist[enumerate]{noitemsep, topsep=3pt, leftmargin=*, align=left}
\setlist[description]{noitemsep, topsep=3pt, leftmargin=*, align=left}

\usepackage{etoolbox}
\patchcmd{\thebibliography}
  {\settowidth}
  {\setlength{\itemsep}{0pt plus 0.1pt}\settowidth}
  {}{}
\apptocmd{\thebibliography}
  {\small}
  {}{}

\usepackage{varwidth}
\usepackage{caption}
\usepackage[percent]{overpic}
\usepackage{subcaption}
\usepackage{amsmath}
\usepackage{cite}
\usepackage{wrapfig}
\usepackage{booktabs}
\usepackage{siunitx}
\DeclareBinaryPrefix\Kilo{K}{10}

\usetikzlibrary{patterns,}
\newcommand{\vars}{\textrm}
\newcommand{\func}{\textsf}
\let\oldReturn\Return
\renewcommand{\Return}{\State\oldReturn}

\algnewcommand{\LineComment}[1]{\Statex  \(\triangleright\) #1 \hfill~}
\usepackage{authblk}

\newcommand*\circled[1]{\footnotesize \tikz[baseline=(char.base)]{
            \node[shape=circle,draw,inner sep=2pt] (char) {#1};}}
            
\everypar{\looseness=-1}

%% file: introduction.tex
\section{Introduction}

Online large-scale data intensive services are becoming increasingly susceptible and sensitive to the modern non-deterministic servers. These type of services are typically fan-out services (i.e., distributed across multiple servers) and  need to act in tandem to ensure user experience.  Prior research has shown that marginal delays (of hundreds of milliseconds) in user experience can greatly impact advertising revenue~\cite{Eric2009TheSearch}. For this reason, user experience is typically expressed in distributed production systems as percentiles of tail latency, such as 95th or 99th percentile. Ensuring consistent tail latency is a hard problem because traditional IPC-based (Instructions per Cycle) scheduling mechanisms have failed~\cite{Lo2015Heracles}.

Given the stochastic nature of the incoming requests to such services, it is important the serve the incoming requests based on their heterogeneous computational demands. One such avenue recently explored in the literature~\cite{Petrucci2015Octopus-Man:Computers,NishtalaHipster,180148,Zhou:2016:GLM:2925426.2926272} to exploit such fan-out services has been heterogeneous multi-core architectures.

Heterogeneous multi-core architectures include types of cores having different power and performance characteristics, typically big/high-performance and little/power-efficient cores sharing the same ISA (Instruction Set Architecture). These heterogeneous multi-cores have emerged as an architecture design to keep meeting performance and energy efficiency requirements for a wide range of applications. The key idea is that threads with high computational load are more suitable to run on big cores, while threads requiring low computational resources can execute more on little cores, leaving space to run additional threads on the big cores.

Modern data centres typically host services such as web search or social network that require strict quality-of-service (QoS) levels~\cite{Kasture2015Rubik,Lo2015Heracles}.  Web search is a critical and challenging workload in data centres due to several factors: the service imposes strict query response time constraints, the arrival time of a query is hard to predict, and each query may have distinct lengths and computing requirements. Prior research~\cite{180148} has shown that online interactive workloads can take advantage of heterogeneous cores and can deliver higher throughput in contrast to homogeneous cores. 

In the data centre domain, an important problem with heterogeneous multi-cores is to decide how best to map the application workload on the most appropriate core type, given the different latency requirements~\cite{JanapaReddi2010WebCores}. Prior work such as Hipster~\cite{NishtalaHipster} and Octopus-Man~\cite{Petrucci2015Octopus-Man:Computers} have shown that heterogeneous cores are attractive to execute cloud workloads, enabling mapping the application at runtime on a set of big or little cores for improved energy efficiency.

By contrast to most recent prior work~\cite{Petrucci2015Octopus-Man:Computers,NishtalaHipster} that map the entire application on heterogeneous cores, this work presents a thread/request-level mapping policy called \textbf{Hurry-up}. Hurry-up can select and execute individual threads on  heterogeneous cores. These threads are typically compute-intensive and are responsible for processing incoming user requests. We show that such a fine-grained solution for thread mapping can lead to improved throughput and tail latency in heterogeneous server systems. 

The main contributions of our work are:

{\footnotesize \circled{1}} We introduce a thread mapping solution, \textbf{Hurry-up}, that exploits a key insight that user queries translate to different computing requirements, such as by varying length of keywords; thus, request threads can be individually mapped at runtime to the most appropriate heterogeneous core for improved tail latency.

{\footnotesize \circled{2}} We implement and deploy Hurry-up on a ARM 64-bit big.LITTLE architecture (Juno R1~\cite{ARMR1https://goo.gl/EcamOa}). We instrument Hurry-up to efficiently manage the execution of threads for a representative Web Search benchmark~\cite{Elasticsearchhttps}.

{\footnotesize \circled{3}} We perform real system evaluation comparing Hurry-up with baseline Linux scheduling (a conservative policy) and show that, Hurry-up improves throughput by XX\% (mean) and reduces tail latency by 39.5\% (mean).

%% file: motivation.tex
\section{Motivation}
\label{sec: motivation}

\begin{figure}[t] 
\centering
\includegraphics[width=0.48\textwidth]{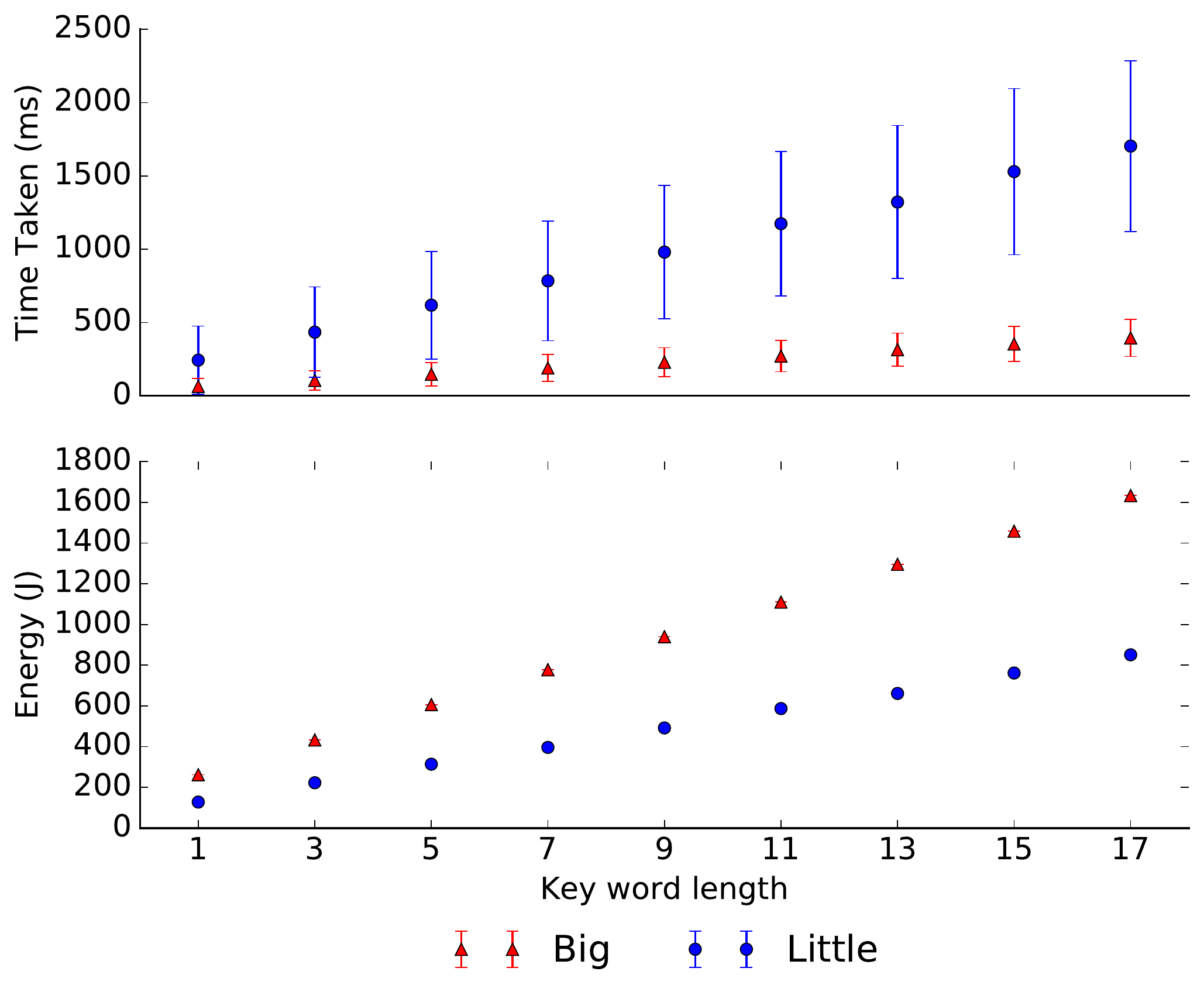}
\caption{Time taken for query processing and energy consumed (in J) when varying the number of keywords while running Web Search on big and little core. The error bars represent the standard deviation.} 
\label{fig:timetaken}
\end{figure}

We motivate our work by performing exploratory experiments on a big/little multi-core architecture running a Web Search benchmark configured to receive different user queries. Details on the experimental setup are given in Section~\ref{subsec: method}.

Figure~\ref{fig:timetaken} shows the processing time required (top plot) and energy consumed (bottom plot, in Joules) for Web Search as a function of big core or little core allocation and query compute-intensity in number of keywords submitted by the users (total of \SI{1e5}{} of requests). The error bars represent the standard deviation for query processing time and energy. If we consider a QoS target at \SI{500}{\milli\second}, user requests with five or more keywords (heavy requests) would violate the latency constraint when running on a little core. On the other hand, requests with fewer than five keywords (light requests) could run on a little core without much impact on the latency target. Note that this is non trivial since these requests experience a lot of variability when running on little cores. In addition, observe the big core can satisfy requests with up to 17 keywords without violating the latency, but consumes higher energy. For all light requests, considering only socket power, we observe that the little core is much more energy efficient (performance per watt of power consumed) than the big core.

\begin{figure}[t] 
\centering
\begin{overpic}[width=0.5\textwidth]{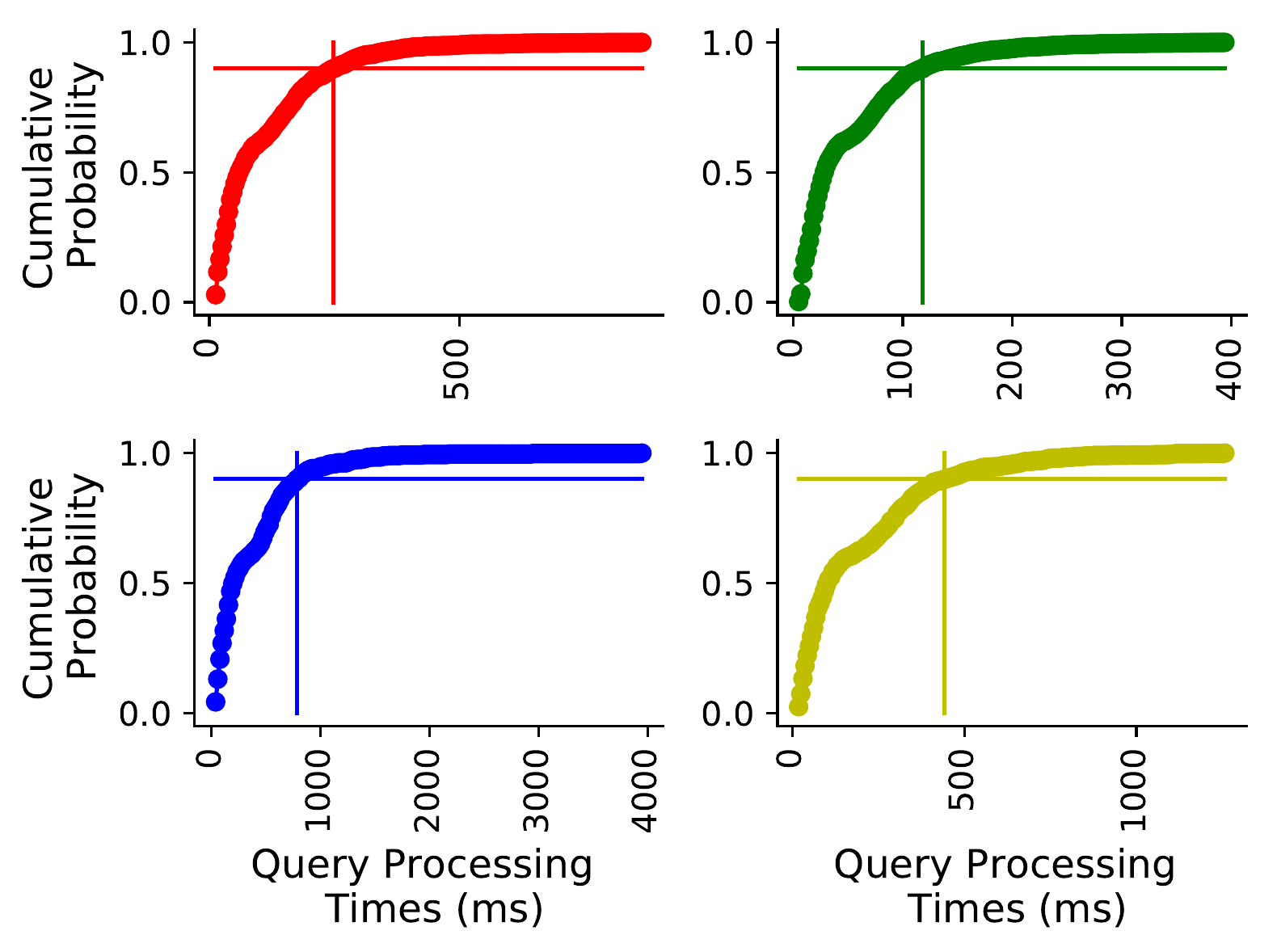}
	\put(27,60){\small Single big core}
      \put(27,25){\small Single little core}
      \put(74,60){\small Two big cores}
      \put(74,25){\small Two little cores}
\end{overpic}
\caption{Query latency distribution on different number of cores (1 or 2) and types (big or little).} 
\label{fig:timetakencdf}
\end{figure}
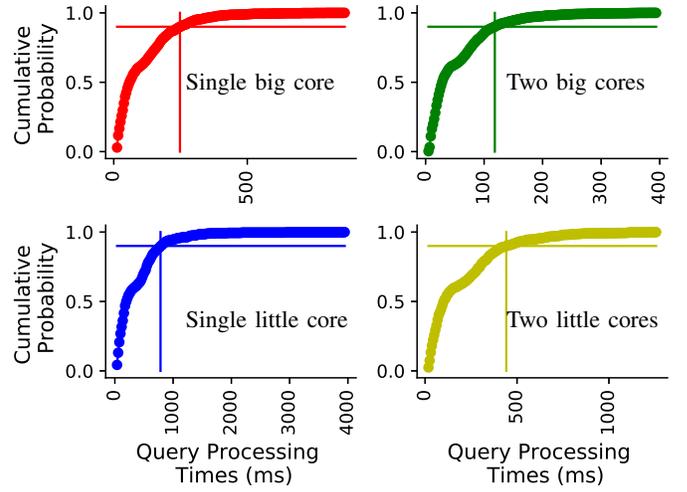

\begin{figure}[t]
\centering 
\begin{tikzpicture}
    \begin{axis}[
            ybar,
            bar width=.8cm,
            width=0.5\textwidth,
            height=.4\textwidth,
            legend style={at={(0.5,1)},
                anchor=north,legend columns=-1},
            symbolic x coords={1-L, 2-L, 1-B, 2-B},
            xtick=data,
            %nodes near coords,
            nodes near coords align={vertical},
            ymin=0,ymax=15,
            ylabel={Normalized to a single little core},
            xlabel={Core Type},
        ]
        \addplot table[x=type,y=Latency,pattern=north east lines]{\latdata}; 
        \addplot table[x=type,y=power]{\latdata};
        \legend{Tail Latency, Socket Power}
    \end{axis}
\end{tikzpicture}
\caption{The tail latency (higher is better) and socket power normalized (lower is better) to a single small core. $B$ and $L$ represent big and little cores, respectively.}
\label{fig: powerstats} 
\end{figure}
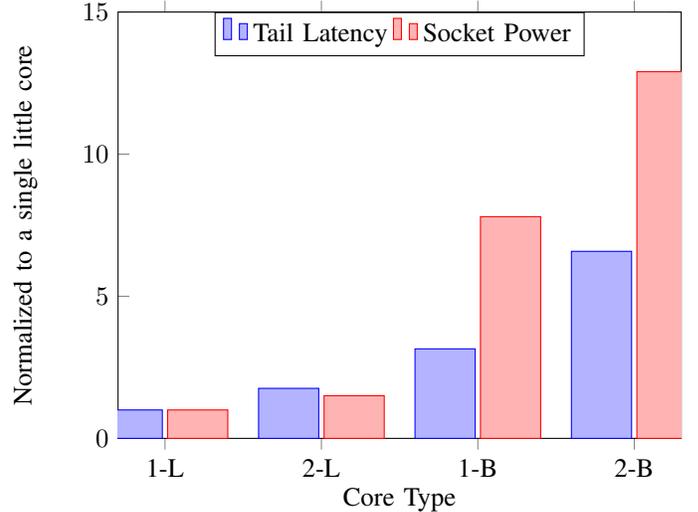

Figure~\ref{fig:timetakencdf} shows an experiment varying the number of cores and core types to see the impact on the tail latency distribution (e.g., 90\%-ile). Considering a QoS target of 90\%-ile at \SI{500}{\milli\second}, the system cannot meet this constraint by using single little core, but it can using two little cores. We note that using one or two big cores, we can greatly reduce the tail latency, but at the expense of using more power. For instance, Figure~\ref{fig: powerstats} represents the socket power consumption and tail latency normalised to a single little core (1-L). The configurations in the $X$-axis, $B$ and $L$, are read as big and little cores, respectively. For tail latency, higher is better whereas for socket power consumption, lower is better. Observe that using single big core can reduce tail latency by up to 3.2$\times$, but consumes 7.8$\times$ higher power in contrast to a single small core. This shows that requests can also experience highly unpredictable variability depending on the core type they will run.

\pgfplotstableread[row sep=\\,col sep=&]{
    type & Latency & power \\
    1-L    &	1	& 1 \\
    2-L & 1.76	&1.5 \\
	1-B	     & 3.15 &	7.8 \\ 
	2-B	 & 6.58 &	12.9 \\
}\latdata

In real server systems, it is impractical to annotate all applications to pass to the scheduler information, such as the number of keywords, that is relevant to compute time. We notice that heavy requests will stay much longer in the system compared to light requests, so we can monitor the progress of each request by reading runtime statistics from the application-level and infer the query computational intensity at runtime.

This motivates the need for a solution that can determine at runtime how to best map a search thread (either light or heavy) on the most suitable core type (either little or big). It is also important that this can be accomplished with minimal additional power expense and thread mapping overhead. 

%% file: algorithm.tex
\section{Hurry-up Thread Mapping Design}

We design Hurry-up to influence the OS scheduling by mapping the threads to run on the core type that can make best usage of the available resources and deliver improved tail latency. 
To make thread mapping decisions, Hurry-up takes advantage of application-level information regarding the execution of each thread in the system. There are two major phases in the Hurry-up design: profiling and mapping.

\subsection{Profiling Phase}

Hurry-up works by firstly identifying specific methods of the application that are suitable for acceleration in the big cores; those are refereed as ``hot functions''. We select via profiling the hottest function present in the critical path a single request execution. We insert monitoring probes via code instrumentation to record the events of entry and exit for such a critical method. In this work, we consider a single hottest function per application in the design.

\subsection{Mapping Phase}

\begin{figure}[t] 
\centering
\includegraphics[width=0.46\textwidth]{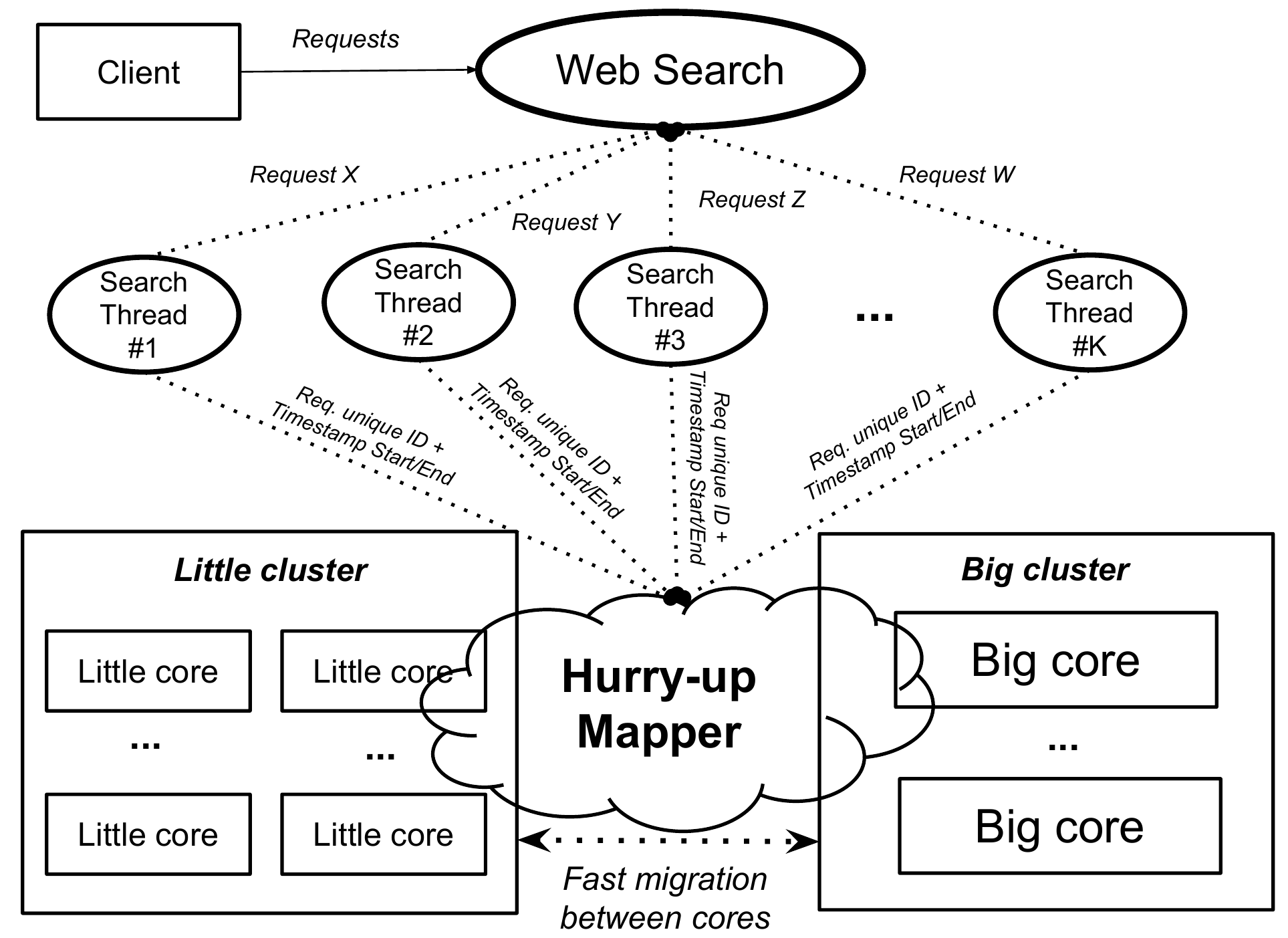}
\caption{Hurry-up Thread Mapping for Web Search} 
\label{fig:diagram}
\end{figure}

As depicted in Figure~\ref{fig:diagram}, a client sends requests to be processed by the Web Search back-end. Each request is internally mapped to a search thread selected from a pool of threads. Once the search thread starts processing the request, it records a timestamp and a unique ID for that particular request. When a search thread finishes processing a request, it also records a timestamp for this event as well.

Hurry-up Mapper reads from a fast communication channel (IPC, Inter-process Communication) the statistics of Web Search application about thread and request identification, and associated timestamps of start/end request processing. Leveraging this runtime observation, collected and updated periodically, Hurry-up Mapper is able to perform thread mapping decisions to improving tail latency, as described next.

\begin{algorithm*}[htbp]
  \caption{Hurry-up Mapper}
  \label{pseudo: algo}
  \begin{algorithmic}[1]      	
     \State $\vars{RequestTable}$ = \{\}
 \State $\vars{StartSamplingTime}$ = \func{GetTimeInMilliSeconds()}
\While {True}
   	\LineComment{Read new request stats: Thread (running) ID, Request (unique) ID, Request (begin/end) Timestamp}
        \State $\vars{TID}$, $\vars{RID}$, $\vars{RTS}$ = \func{ReadStatsFromApp()} 
        \If{$\vars{RID}$ in $\vars{RequestTable}$}
     	\State \textbf{delete} $\vars{RequestTable}[\vars{RID}]$ \Comment{Request already finished}
    	\Else
    	\State $\vars{RequestTable}[\vars{RID}]$ = ($\vars{TID}$, $\vars{RTS}$) \Comment{Store new request data}
    	\EndIf
        \If{($\func{GetTimeInMilliSeconds()} - \vars{StartSamplingTime}$) $<$ $\vars{SAMPLING\_TIME}$}
     	\State \textbf{continue} \Comment{Restart loop to keep reading more data}
    	\EndIf
        \State $\vars{ThreadsOnLittle}$ = []
        \For{($\vars{TID}$, $\vars{RTS}$) in $\vars{RequestTable}$}
        \State $\vars{TimeElapsed}$ = \func{TimeinMilliSeconds()} - $\vars{RTS}$
        \If{($\vars{TimeElapsed}$ $>$ $\vars{MIGRATION\_THRESHOLD}$)}
        \If{$\vars{TID}$ is running on little core type}
     	\State Add pair ($\vars{TID}$, $\vars{TimeElapsed}$) on $\vars{ThreadsOnLittle}$ 
        \EndIf
    	\EndIf
     \EndFor
      \State Sort $\vars{ThreadsOnLittle}$ by $\vars{TimeElapsed}$ in descending order
      \For{$\vars{b}$ = 0$\,\to\,$ $\func{size}(\vars{BigCoreList}$)}
        \If{$\vars{b}$ $>=$ $\func{size}(\vars{ThreadsOnLittle}$)} 
     	\State \textbf{break} \Comment{No more migration requests on little cores}
        \EndIf
         \State $\vars{BigCore} = \func{BigCoreList}[\vars{b}]$
         \State $\vars{ThreadOnBig} = \func{GetRunningThread}(\vars{BigCore})$
         \State $\vars{ThreadID} = \vars{ThreadsOnLittle}[\vars{b}]$
         \State $\vars{LittleCore} = \func{GetRunningCore}(\vars{ThreadID})$
         \State Map $\vars{ThreadID}$ to $\vars{BigCore}$
         \State Map $\vars{ThreadOnBig}$ to $\vars{LittleCore}$
     \EndFor		      
     \State $\vars{StartSamplingTime} = \func{GetTimeInMilliSeconds()} $   
  	\EndWhile    
  \end{algorithmic}
\end{algorithm*}

\subsection{Hurry-up Mapper}

A high level description of Hurry-up Mapper is described in Algorithm~\ref{pseudo: algo}. Given the incoming user requests, Hurry-up is responsible for mapping the search thread serving the request to either a big or little core.
Hurry-up works by mapping a light search thread that can potentially finish its execution on a little core without much impact on the tail latency, and a heavy search thread that is more compute-intensive on a big core to improve tail latency.

The empirically tuned parameters for the algorithm are: \texttt{SAMPLING\_TIME} that controls how frequently we sample runtime statistics from the application and \texttt{MIGRATION\_THRESHOLD} that specifies a time threshold used to identify a thread as compute-intensive and migrate the thread to a big core.

Each search thread has unique ID in the pool and is responsible for processing a request (also having unique ID); a thread needs to finish an active request before another request can be processed~\cite{Elasticsearchhttps}. The initial mapping of the search thread pool is carried out in a round-robin fashion so that the workload is balanced among all the available cores uniformly.

In Algorithm~\ref{pseudo: algo}, Lines~1-2 initialise \textit{RequestTable} for storing the runtime data and \textit{StartSamplingTime} to determine the start time of a sampling window. Line~4 is responsible for collecting statistics from the application in the form of \textit{timestamps} of each thread-request unique pair. Next, Lines~5-8 check if it is a new request or a request that has finished its processing. All finished requests are removed from \textit{RequestTable}. Lines~9-10 control the loop of runtime data reading. 

The \textit{ReadStatsFromApp} function (line 4) reads the application data from a pipe channel (interprocess communication). It blocks waiting in case there is no available data. An example of a stream of data obtained after a call of \textit{ReadStatsFromApp} is shown below:

\small{
\begin{verbatim}
75;ixI.;1498060927539
77;1J.D;1498060927953
78;579[;1498060927954
79;Xrt@;1498060928003
80;qc8o;1498060928014
77;1J.D;1498060928023
\end{verbatim}
}

Analysing the snapshot above at a given sampling interval, the first line indicates that Thread ID $75$ with request ID \texttt{ixI.} started at timestamp $1498060927539$, but it is still in progress because looking further in the available data there is no event indicating the end of its execution. Next, Thread ID 77 (request ID \texttt{1J.D}) started at timestamp $1498060927953$; as shown further in the last line, it finished at timestamp $1498060928023$ and took 70 ms of execution time (subtraction of $1498060928023-1498060927953$). With the given data, we note that all other Thread IDs such as 78, 79 and 80 are still processing their requests.

%\pmc{system-on-chip?}\nishtala{what do you mean?}
Lines~11-16 use the request table to identify any request-thread pair (to append in \textit{ThreadsOnLittle} list) that has been long running on a little core for at least the migration time threshold (in \SI{}{\milli\second}). In Line~18, the threads running on little cores are sorted in the descending order of their time elapsed in the system. Finally, Lines~18-26 are responsible for the actual remapping of the threads on little cores. Each long running thread on a little core (starting from the longest thread) is selected to run on a big core, until there are no more big cores or no more migrative threads on little cores. Line~27 reset the start sampling time and the algorithm loop is resumed.

We empirically set the sampling interval (in \SI{}{\milli\second}) to ensure the algorithm has enough runtime data to make the thread mapping decisions. In our case, we found that \SI{50}{\milli\second} worked best for periodically reading the runtime data with low overhead, while any other longer sampling times performed worse. On the other hand, we can notice that the algorithm is very sensitive to the migration threshold time, because it is responsible for triggering the mapping of threads from little to big cores.
In the next section, we will evaluate how the Hurry-up algorithm performs in a real big/little platform and present the sensitivity analysis.

%% file: evaluation.tex
\section{Experimental Results}
\label{sec:experiments}

\begin{figure}[t] 
\centering
\includegraphics[width=0.35\textwidth]{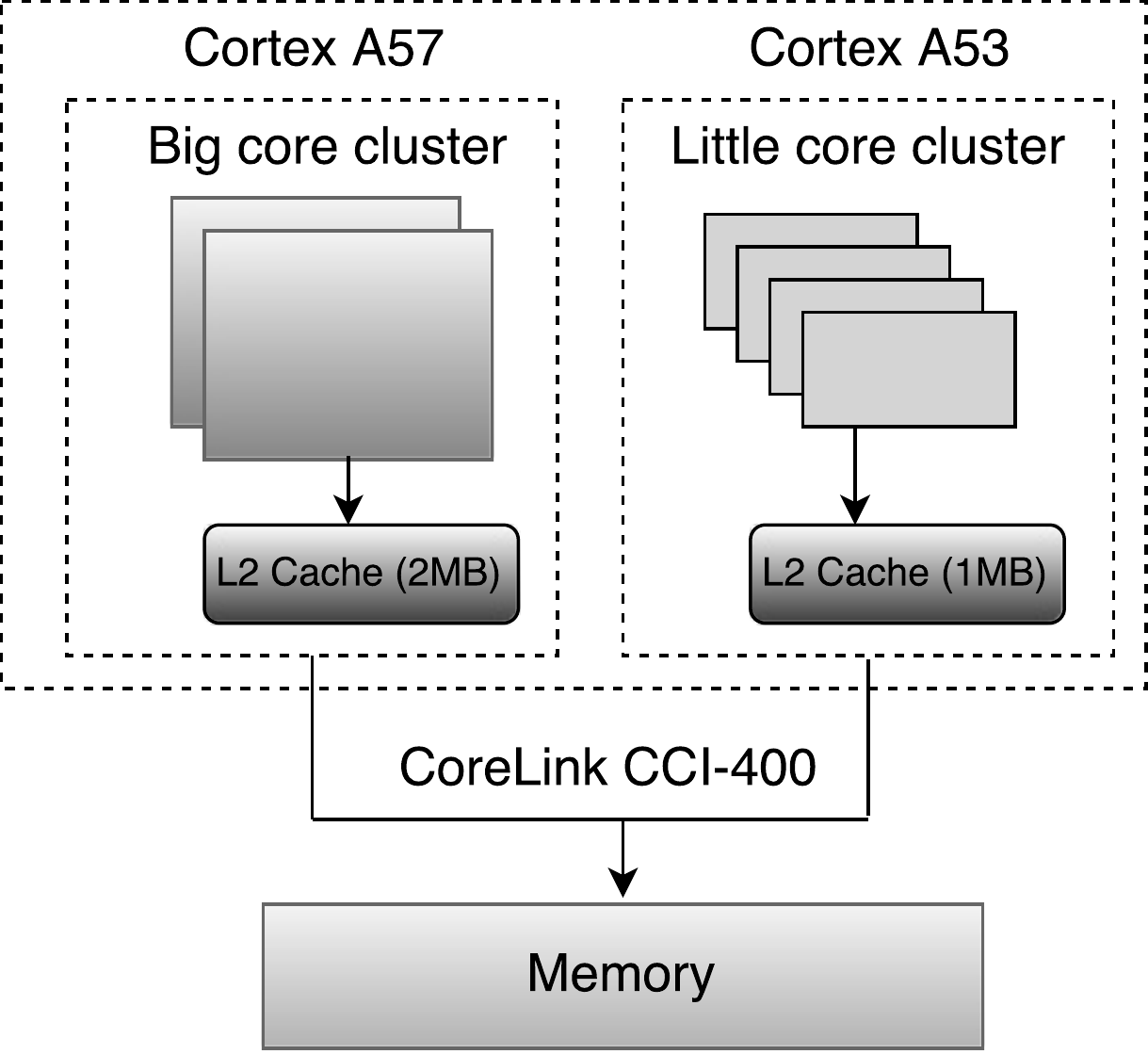}
\caption{Heterogeneous multi-core platform (ARM Juno R1)} 
\label{fig: archiarm}
\end{figure}

\subsection{Methodology}
\label{subsec: method}
\textbf{[Big/Little System]:} We perform the evaluation experiments on an ARM Juno R1 developer board~\cite{ARMR1https://goo.gl/EcamOa} with Linux (kernel 4.3). The Juno board is a 64-bit ARMv8 big.LITTLE architecture with two high-per\-for\-mance out-of-order Cortex-A57 (big) cores and four low-power in-order Cortex-A53 (little) cores. The cores are integrated on a single chip with off-chip \SI{8}{\giga\byte} DRAM. The two big cores form a cluster with a shared \SI{2}{\mega\byte} L2 cache, and the four little cores form another cluster with a shared \SI{1}{\mega\byte} L2 cache. The big and little cores are set to the highest DVFS state of \SI{1.15}{\giga\hertz} and \SI{0.6}{\giga\hertz}, respectively.
The cache interconnect (CoreLink CCI-400) provides full cache coherency among the heterogeneous cores, allowing a shared memory application to run on both clusters simultaneously.

\textbf{[Power Measures]:} The power consumption of the Juno board is obtained using four native energy meters~\cite{ARMRegistershttps://github.com/ARM-software/devlib/blob/master/src/readenergy/readenergy.c}. The four  energy meters are responsible for collecting results from the big cluster, little cluster, rest of the system (including memory controllers, etc) and the Mali GPU.  The power consumption of the Mali GPU is negligible because the GPU is disabled in all our experiments. The system power consumption is reported as an aggregation of the big and little clusters, and the rest of the system (including memory controllers, etc). We observe that a single big core is 52\% more power-efficient than a single little core, in terms of IPS (Instructions per Second) per watt. But, taking into account all cores in a cluster, and assuming that all cores can be fully utilised, a little cluster is 25\% more power-efficient than a big cluster. This discrepancy is because the rest of the system, excluding the core clusters, consumes about the same power as the big core at full utilisation (\SI{0.76}{\watt}). If we subtract the power of the rest of the system, a single little core is $2.3\times$ more power-efficient than a big core. The little cores are attractive to improve the throughput of sequential workloads, due to their power-efficient characteristics. Big cores are, however, still necessary for lowering the tail latency, as a result of computationally-intensive single-threaded requests.

\begin{figure}[t] 
\centering
    \begin{overpic}[width=\linewidth]{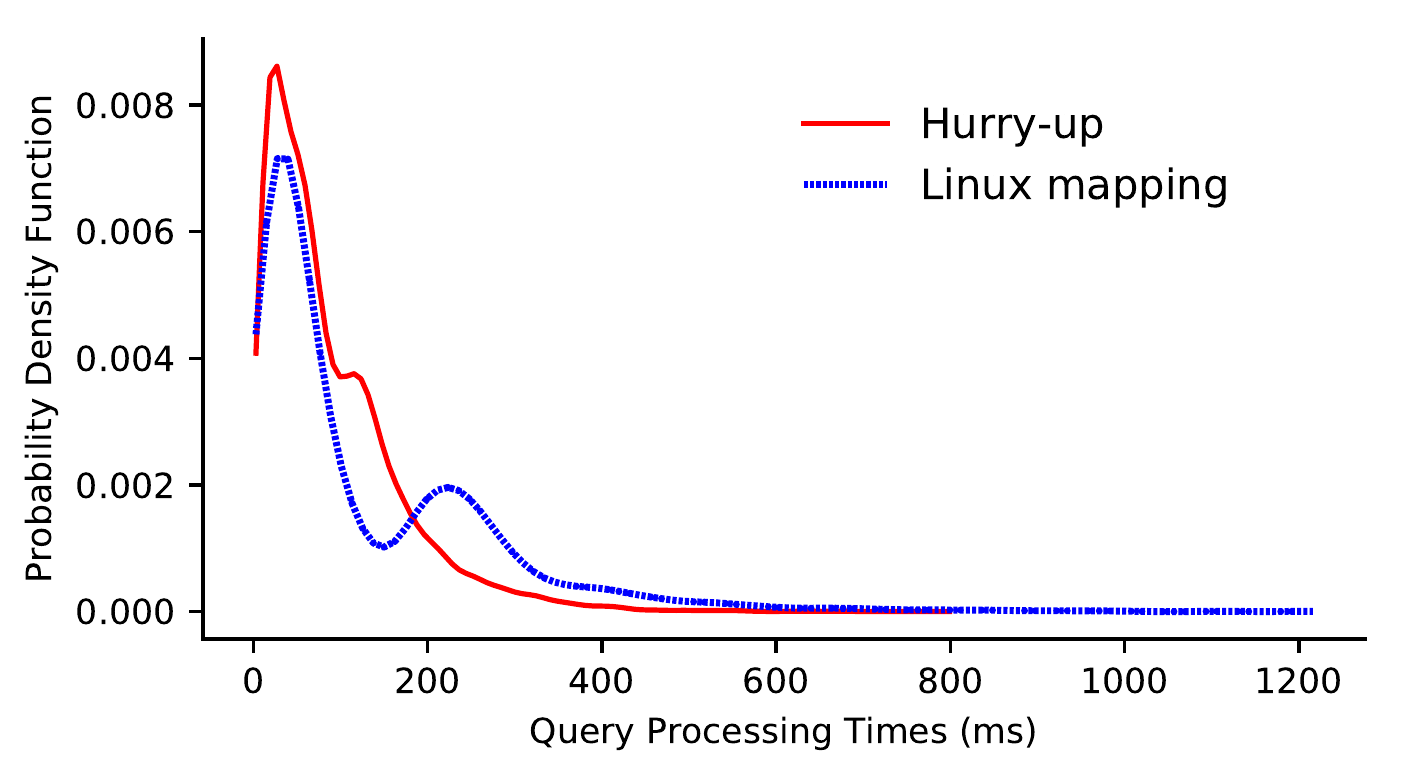}
      \put(25,46){\Large B}
      \put(29,25){\Large C}
      \put(68,13){\Large A}
     \end{overpic}
\caption{Latency distribution: Hurry-up vs Linux mapping.}
\label{fig:pdfdata}
\end{figure}

\textbf{[Benchmark]:} 
We evaluate Hurry-up using a Web Search benchmark. Typically, search engines are designed as scale-out workloads to deliver high request throughput under strict levels of latency constraints~\cite{JanapaReddi2010WebCores,Barroso2003WebArchitectureb}. A single user request is fanned-out to many leaf node servers to process the query on their shard of the search index~\cite{Lo2015Heracles}. We experiment with a big/little architecture server running Elasticsearch~\cite{Elasticsearchhttps}, an open source implementation of a search engine used by many companies including Netflix and Facebook. The Elasticsearch has an index of the English Wikipedia database that fits in the server memory. We configure the size of the search thread pool (six threads) to match the number of cores in the system (two big cores plus four little cores). Unless otherwise stated, we specify the \textbf{tail latency} as the 90th percentile response latency. The load generator (Faban) for Web-Search is adapted from CloudSuite 3.0~\cite{Ferdman2012ClearingClouds}.  The maximum load is chosen such that platform can still improve throughput when running on two big cores and four little cores at maximum DVFS state without impacting too much on the long-tail latency. In our experiments, the load generator simulates the load (i.e., clients) on another machine: an AMD Opteron 6140 64-bit with eight cores at \SI{2.6}{\giga\hertz} and \SI{32}{\giga\byte} DRAM. The load generator machine is connected via \SI{1}{\giga\bit} network to the big/little server machine.

\subsection{Hurry-up Results}

We show the effectiveness of Hurry-up by comparing against  a conservative/static Linux mapping policy. The Linux baseline maps each request to a given core type randomly, and there exists no migrations thereafter. In contrast, Hurry-up works by dynamically mapping individual queries on a heterogeneous platform to improve tail latency.

\textbf{[Tail Latency]:} We show that Hurry-up can reduce the query processing time and maximise the total number queries served. To demonstrate this, we show the Probability Distribution Function (PDF) in Figure~\ref{fig:pdfdata} from a experiment in which we sampled the query processing time for each request with a simulated load of 30 QPS using Faban. For Hurry-up, we set the sampling interval and migration threshold to \SI{25}{\milli\second} and \SI{50}{\milli\second} respectively. In the plot, as can be seen at point A, Hurry-up reduces the worst case tail latency from \SI{1200}{\milli\second} to \SI{800}{\milli\second}. At the other extreme, looking at point B, Hurry-up shows a higher density than that of Linux mapping because Hurry-up aggressively migrates potential, but not certain, long-running requests from little to big cores. At point C, using Hurry-up, we notice that the requests migrated to a big core much earlier compared to Linux mapping. In Linux mapping, the requests continue to execute on little cores, increasing their processing time.

\begin{figure}[tb]
  \centering     
     \includegraphics[width=0.5\textwidth]{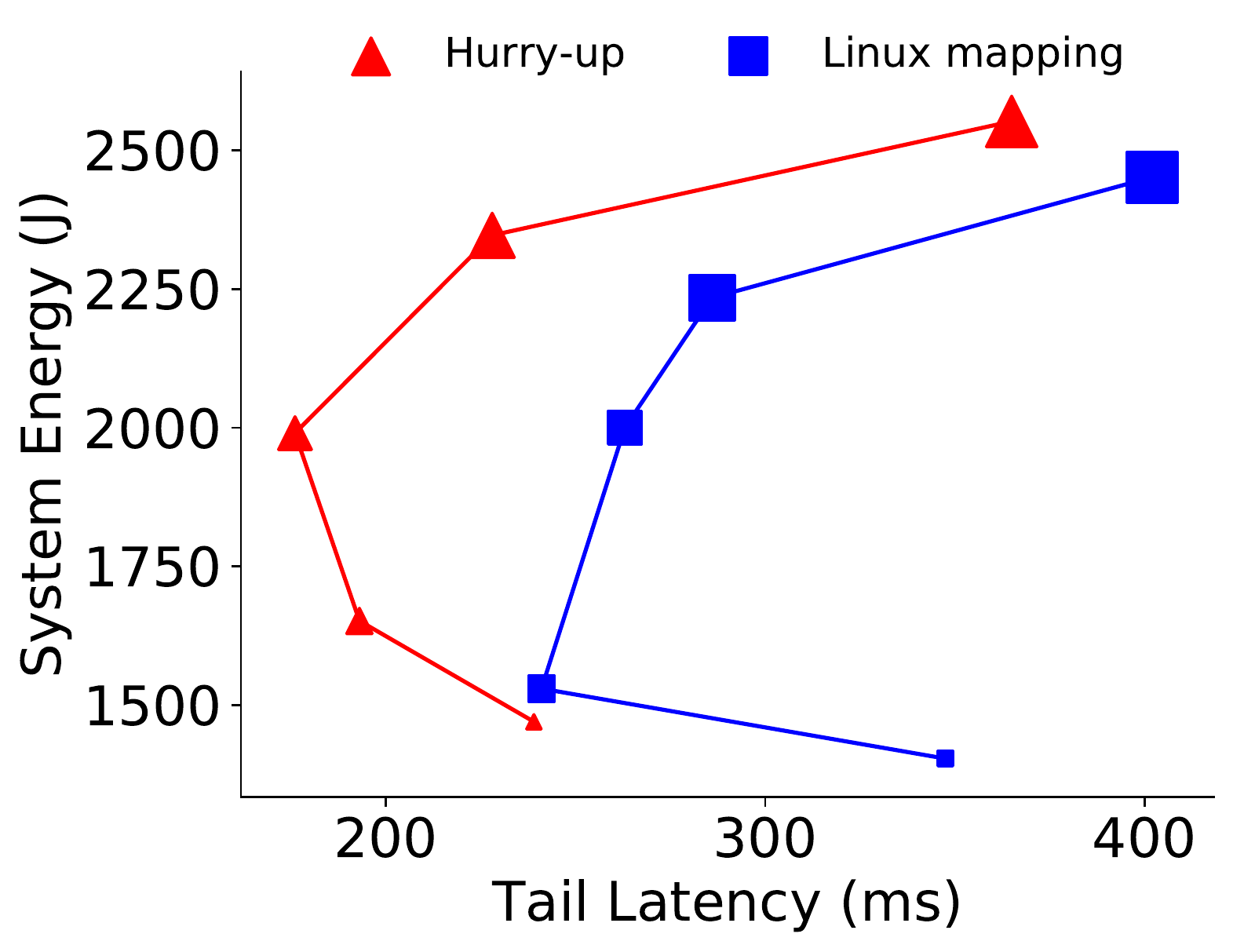}
	\caption{Trade-off between tail latency and system energy (in Joules)  for Hurry-up and Linux mapping. The size of the data point represents the load in QPS (5, 10, 20, 30 and 40).}
\label{fig: QPS-Latency} 
\end{figure}

\textbf{[Tail latency vs. Energy]:} Figure~\ref{fig: QPS-Latency} shows the trade-off between tail latency and system energy consumption for policies Hurry-up and Linux mapping. The size of the scatter point represents the load (i.e., the smallest point represents least load, whereas the largest point represents highest load). For each policy, we conducted an experiment with load fixed at 5, 10, 20, 30 and 40 QPS. We make two observations. 

{\footnotesize \circled{1}} Hurry-up has a lower tail latency in contrast to Linux mapping while having a slightly higher  energy consumption (4.6\% mean). This is because, Hurry-up maps heavy requests from little cores to big cores after a migration threshold and allocates little core to lighter requests. This helps Hurry-up improve the tail latency but increases energy consumption as it utilises the bigger core for a longer duration.  By contrast, Linux mapping (may) execute light requests on the big core, while keeping the big core idle for extended periods and thereby consumes lower energy and a higher tail latency.

{\footnotesize \circled{2}} Observe at low load (5 QPS), Hurry-up has a higher tail latency than at higher loads (10, 20 and 30 QPS) because, at low loads a larger percentage of the requests are executed on little cores in contrast to big core, whereas the contrary is true at high loads. For instance, with 5 QPS there are approximately 33\% of the requests are executed on big cores while 67\% of the requests are executed on little cores. On the other hand, with 20 QPS, approximately 58\% of the requests are executed on big cores and the remaining on the little cores. As the number of requests processed on big cores increase, so does the energy consumption but the tail latency reduces until there exists no queuing.

\pgfplotstableread[row sep=\\,col sep=&]{
    QPS & LH & LS \\
    5	&239	&	347\\
	10	&193	&	241\\
	15	&174	&	252 \\
	20	&176	&	328	\\
	30	&228	&	286	\\
	40	&365	&	402	\\
    }\mydata

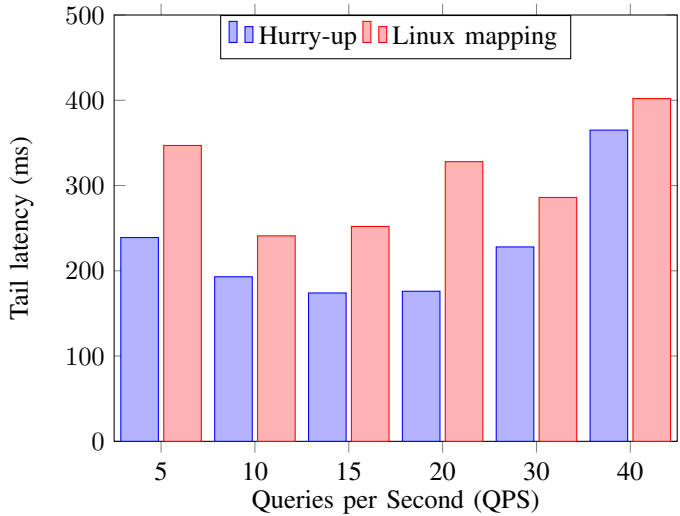
\begin{figure}[t]
\centering 
\begin{tikzpicture}
    \begin{axis}[
            ybar,
            bar width=.5cm,
            width=0.5\textwidth,
            height=.4\textwidth,
            legend style={at={(0.5,1)},
                anchor=north,legend columns=-1},
            symbolic x coords={5, 10, 15, 20, 30, 40},
            xtick=data,
            nodes near coords align={vertical},
            ymin=0,ymax=500,
            ylabel={Tail latency (ms)},
            xlabel={Queries per Second (QPS)},
        ]
        \addplot table[x=QPS,y=LH]{\mydata};
        \addplot table[x=QPS,y=LS]{\mydata};
        \legend{Hurry-up, Linux mapping}
    \end{axis}
\end{tikzpicture}
\caption{The tail latency (in \SI{}{\milli\second}) at various loads (in QPS) for policies Hurry-up and Linux mapping.}
\label{fig: hurry-static} 
\end{figure}

\textbf{[Impact on tail latency]:} Figure~\ref{fig: hurry-static} shows the tail latency (in \SI{}{\milli\second}) at various loads (in QPS) for policies Hurry-up and Linux mapping. For Hurry-up, we set the sampling interval and migration threshold to \SI{25}{\milli\second} and \SI{50}{\milli\second}, respectively.  In the plot, as can be seen, Hurry-up reduces the worst-case tail latency at all loads in contrast to Linux mapping because, it migrates heavy requests from little to big core after a given migration threshold. This allows Hurry-up to reduce tail latency by up to a maximum of 86\% at 20 QPS and by 39.5\% on average. On the other hand, at the highest load of 40 QPS, Hurry-up can only reduce by only 10\% due to the high-tail latency (and queuing) experienced by both scheduling policies.

\begin{figure*}[t] 
\centering
\includegraphics[width=\textwidth]{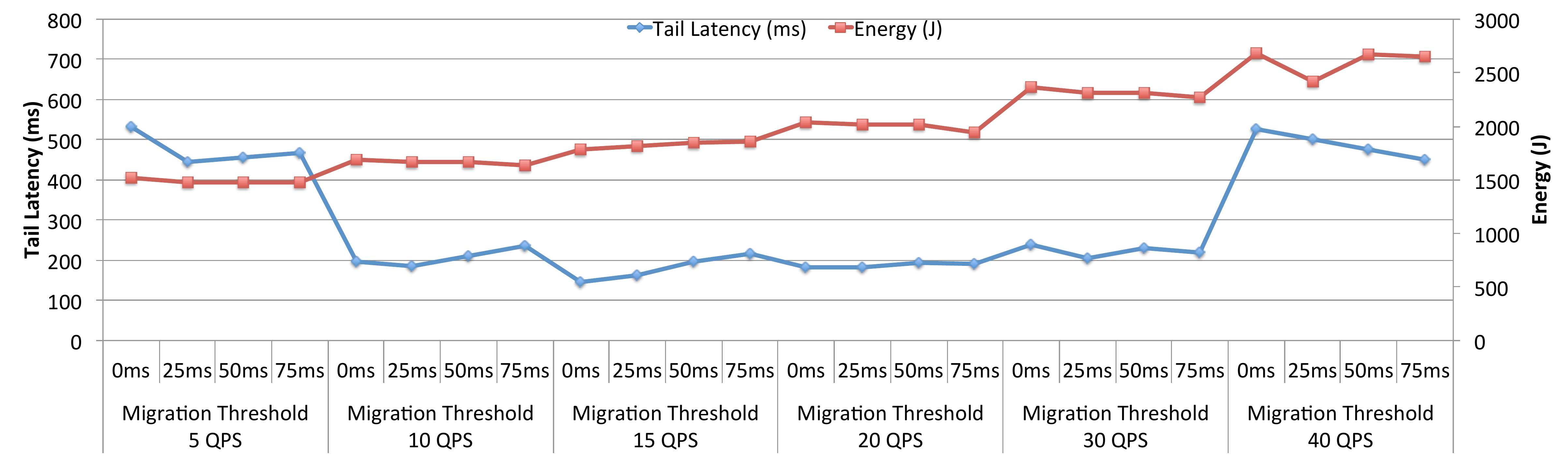}
\caption{Distribution of tail latency and energy (in J) as a function of migration threshold and the load (in QPS), when sampling interval is set to \SI{50}{\milli\second}. The primary $y$-axis shows the tail latency (in \SI{}{\milli\second}) and the secondary $y$-axis shows the energy (in J).}
\label{fig: sensitivity}
\end{figure*}

\textbf{[Parameters sensitivity]:} To make the best use of Hurry-up, we empirically tune the parameters: sampling interval and migration threshold, and select the parameters that deliver the best balance between tail latency and energy. Figure~\ref{fig: sensitivity} shows the distribution of the tail latency and energy as a function of migration threshold and the load (in QPS), when the sampling interval is set to \SI{50}{\milli\second} for policy Hurry-up. We avoid showing for all sampling intervals to avoid visual clutter. The primary $y$-axis shows the tail latency (in \SI{}{\milli\second}) and secondary $y$-axis shows the system energy consumption (in joules). Similar to Figure~\ref{fig: QPS-Latency}, notice that at lowest load (5 QPS) there exists a high-tail latency because fewer requests are completed on big core in contrast to little core; whereas at the highest load (40 QPS), despite having a large number of requests being processed on big cores, the tail latency increases due to queuing. In addition, observe that with load set to 10 QPS, 15 QPS, 20 QPS and 30 QPS, a higher migration threshold entails a higher latency and lower energy consumption. This is because, complex requests running on little cores are migrated to big cores after a longer migration threshold and thereby executing requests longer on little cores and consuming lower power. On the contrary, a lower migration threshold entail a lower latency and higher energy consumption as both, simple and complex requests are migrated to big cores quicker and thereby consuming higher power. 

%% file: relatedwork.tex
\section{Related Work}
\label{section: RelatedWork}

Twig~\cite{NishtalaTwig} introduced a deep reinforcement learning based solution to manage cores and DVFS control for multiple latency-critical workloads to improve energy efficiency. 
Hipster~\cite{NishtalaHipster} introduced a hybrid scheme that combines heuristics and reinforcement learning to manage heterogeneous cores with DVFS control for improved energy efficiency and resource utilisation. 
Octopus-Man\cite{Petrucci2015Octopus-Man:Computers} was designed for big.LITTLE architectures to map workloads on big and little cores using a feedback controller in response to changes in measured latency. 
Adrenaline~\cite{Hsu2017adrenaline} uses application level hints to identify heavy threads that can affect the tail latency, and provides a scheme for boosting those queries exploiting quick frequency/voltage scaling. 
Ren~\etal~\cite{180148} investigate workloads that maximise throughput on heterogeneous processors, and demonstrate that heterogeneous processors deliver up to 50\% higher throughput in contrast to homogeneous cores. GreenGear~\cite{Zhou:2016:GLM:2925426.2926272} proposed a heterogeneous platform-aware power provisioning system for data centres. 
The management framework distributes power from either renewable and non-renewable sources between little and big cores to achieve a higher energy efficiency while meeting SLO targets. KnightShift~\cite{Wong2012KnightShift:Heterogeneity} introduces a server architecture that couples commercial available compute nodes to adapt the changes in system load and improve energy proportionality (i.e., system power consumption is proportional to utilisation).

Haque~\etal~\cite{Haque:2015:FIP:2775054.2694384} introduce a few-to-many parallelism technique that dynamically increases request-level parallelism at runtime. Their system completes simple/less complex requests sequentially to save resources, and parallelizes larger requests to reduce tail latency. 
Kim~\etal~\cite{Kim:2015:DPR:2684822.2685289} estimate the tail latency of each request using machine learning to execute the estimated request sequentially to save resources, and to parallelise larger requests to reduce latency slack. Li~\etal~\cite{Li:2016:WSI:3016078.2851151} proposed an approach to improve service level objectives (SLO) at request-level. They serialised the complex requests on the system to reduce the impact of queuing on less complex requests.

Heracles~\cite{Lo2015Heracles} uses a feedback controller that exploits collocation of latency-critical and batch workloads while increasing the resource efficiency of CPU, memory and network as long as QoS target is met. Pegasus~\cite{Lo2014TowardsWorkloads} achieves high CPU energy proportionality for low latency workloads using fine-grained DVFS techniques. Time Trader~\cite{Vamanan2015TimeTrader:Search} and Rubik~\cite{Kasture2015Rubik} exploit request queuing latency variation and apply any available slack from queuing delay to throughput-oriented workloads to improve energy efficiency. Quasar\cite{Delimitrou2014Quasar} use runtime classification to predict interference and collocate workloads to minimise interference. 

Mars~\etal~\cite{Mars:2011:BIU:2155620.2155650,Yang2013Bubble-flux} detect at runtime the memory pressure and find the best collocation to avoid negative interference with latency-critical workloads. They also have a mechanism to detect negative interference allocations via execution modulation. 

%% file: conclusion.tex
\section{Conclusion}
\label{section: conclusion}

This paper presented Hurry-up, a thread mapping approach that optimises for tail latency and improves throughput by accelerating long-running compute-intensive requests on big cores, while letting light and less intensive threads running on little cores. Hurry-up recognises that search queries can require different compute requirements, and such a knowledge can be inferred at runtime based on application-level statistics. We show that Hurry-up outperforms a conservative policy under Linux in terms of reducing tail latency by 39.5\% (mean), while requiring negligible additional energy.